\documentclass[sigconf,10pt]{acmart}
\pdfoutput=1
\usepackage{epstopdf}
\usepackage[english]{babel}
\usepackage{blindtext}
\usepackage{color}
\usepackage{multirow}
\usepackage{amsmath}
\usepackage{balance}
\usepackage[colorinlistoftodos]{todonotes}
\usepackage{soul} 
\usepackage{subcaption}
\usepackage{ulem}
\usepackage{algorithm2e}
\setcopyright{none}

\usepackage{enumitem}
\newcommand{\squishlist}{\begin{itemize}[itemsep=1pt,parsep=2pt,topsep=3pt,partopsep=0pt,leftmargin=0em, itemindent=1em,labelwidth=1em,labelsep=0.5em]}
\newcommand{\squishend}{\end{itemize}}
\newcommand{\squishenum}{\begin{enumerate}[itemsep=1pt,parsep=2pt,topsep=3pt,partopsep=0pt,leftmargin=0em,listparindent=1.5em,labelwidth=1em,labelsep=0.5em]}
\newcommand{\squishsubenum}{\begin{enumerate}[itemsep=1pt,parsep=2pt,topsep=0pt,partopsep=0pt,leftmargin=0em,listparindent=1.5em,labelwidth=1em,labelsep=0.5em]}
\newcommand{\squishenumend}{\end{enumerate}}

\copyrightyear{2019} 
\acmYear{2019} 
\acmConference[MobiCom '19]{The 25th Annual International Conference on Mobile Computing and Networking}{October 21--25, 2019}{Los Cabos, Mexico}
\acmBooktitle{The 25th Annual International Conference on Mobile Computing and Networking (MobiCom '19), October 21--25, 2019, Los Cabos, Mexico}
\acmPrice{}
\acmDOI{}
\acmISBN{}

\newcommand{\name}{living IoT}
\newcommand{\Name}{Living IoT}

\newfont{\coprimary}{phvr8t at 10pt}

\begin{document}
\fancyhead{}
\settopmatter{authorsperrow=0}
\title{Living IoT: A Flying Wireless Platform on Live Insects}


\author{Vikram Iyer*, Rajalakshmi Nandakumar*\\
Anran Wang*, Sawyer B. Fuller, Shyamnath Gollakota}
\affiliation{{*Co-primary Student Authors}}
\email{{vsiyer, rajaln, anranw, minster, gshyam}@uw.edu}
\affiliation{%
  \institution{University of Washington}
}


\newcommand{\xref}[1]{\S\ref{#1}}
\newcommand{\red}[1]{\textcolor{red}{#1}}
\newcommand{\del}[1]{\sout{#1}}
\begin{abstract}
Sensor networks with devices capable of moving could enable applications ranging from precision irrigation to environmental sensing.  Using mechanical drones to move sensors, however, severely limits operation time since flight time is limited by the energy density of current battery technology. We explore an alternative, biology-based solution: integrate sensing, computing and communication functionalities onto live flying insects  to create a mobile IoT platform.

Such an approach takes advantage of these tiny, highly efficient biological insects which are ubiquitous in many outdoor ecosystems, to essentially provide mobility for free. Doing so however requires addressing key technical challenges of power, size, weight and self-localization in order for the insects to perform location-dependent sensing operations as they carry our IoT payload through the environment. We develop and deploy our platform on bumblebees  which includes backscatter communication, low-power self-localization hardware, sensors, and a power source. We show that our platform is capable of  sensing, backscattering data at 1~kbps when the insects are back at the hive,  and localizing itself up to distances of 80~m from the access points, all within a total weight budget of 102~mg.
\end{abstract}



\maketitle

\section{Introduction}

Mobility in sensor networks has the potential to transform agriculture by enabling smart farming applications including precision irrigation~\cite{irrigation} and environmental sensing~\cite{environ}.  Sensor mobility significantly reduces the overhead of manual sensor deployment and upkeep, which remains a major barrier to adoption of smart farms~\cite{usagriculture}. Drones, which have thus far been the platform of choice for enabling mobility, are severely power constrained and  last for only 5--30 minutes on a single charge due to the energy density limits of existing battery technologies~\cite{batterylimit}. This is mainly because  the motors drones use for mechanical propulsion and control are power consuming and unlike digital electronics, cannot scale with Moore's law. As a result, the majority of a drone's power consumption is in its propulsion and control systems~\cite{Escobar-Alvarez2018}.


\begin{figure}[t!]
\centering
\begin{subfigure}{0.22\textwidth}
    	\centering
    	\includegraphics[width=1\textwidth]{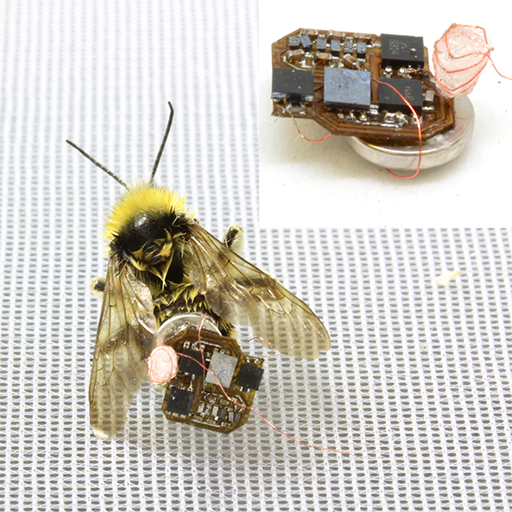}
	\end{subfigure}
    \hspace{1em}
    \begin{subfigure}{0.22\textwidth}
    	\centering
    	\includegraphics[width=1\textwidth]{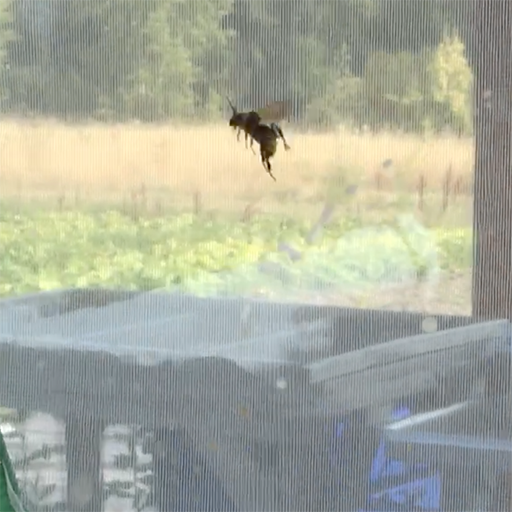}
	\end{subfigure}
  \vskip -1em
  \caption{\small{{\Name\ platform on a bumblebee.} (Left) Bee carrying our platform; (Right) Bee flying with our platform.}}
  \label{fig:bee1}
\end{figure}

This paper explores the idea of creating mobile wireless sensors by placing them on live insects. Using live insects such as bees is attractive for two key reasons. 
\squishlist

\item {\bf Flight Time.}  Unlike drones, flying  insects use chemical energy stored in fats and carbohydrates, which have a much higher energy mass-density than batteries. This allows for much longer flight times: flies have been shown to fly for hours without food~\cite{Dethier1976}, while worker bees spend most daylight hours foraging for nectar and pollen~\cite{foraging_time,foraging_trip_duration} and can fly while carrying payloads of over 100~mg~\cite{beeradio}. Further, these animals have evolved  to have aerodynamic and musculoskeletal systems that minimize  power usage~\cite{Dudley2002,Dickinson1999}. 

\item {\bf Ubiquity.} Insects are nearly ubiquitous across the planet {and adapted to live in diverse ecosystems, making it easy to find a species well suited for a particular environment or application.} Moreover, while some are regarded as pests, others are essential to human activities. For example bees are needed to pollinate many commercial crops, and are in many case intentionally introduced for that purpose~\cite{koppert}.

\squishend

{We piggyback on these insects} to enable mobility for sensor networking applications including  smart farms. Using live insects like bees however introduces two key challenges: 1) they are physically small and can only carry small payloads which severely limits options for power, computation and communication, and 2)  we  cannot easily control the flight of small insects like bees (see~\xref{sec:related}). 	

We present {\it\Name}, a novel {general-purpose wireless sensing platform} that is low-power, low-weight and can support computing and communication operations on flying insects like bees. In order to meet their stringent size and weight requirements, we present a backscatter based communication system that can be achieved with commercially available microcontrollers and has a small and light-weight form factor compatible with insects. Our design includes the capability to compute as well as sample onboard sensors. Additionally backscatter offloads many power expensive components in radios to an  access-point (AP) set up near the hive, enabling low power operations.

To address the lack of flight control, our insects localize themselves in the 2D space using RF signals transmitted by access points in the environment. This self-localization architecture,  similar in spirit to GPS, is attractive because it does not require the AP to estimate the locations and send them back to each insect. Thus, it can scale well with the number of insects, works at high speeds and does not require the insects to transmit signals. More importantly, this enables location-based mobile sensors where the insect can associate location information with its sensing data as well as perform sensing operation only when they pass over the desired set of locations. The  sensor data can then be uploaded to the AP when it returns within range of the hive.

Achieving self-localization on our \name\ platform is however challenging for three key reasons: \squishlist

\item It should be low-power in nature as we cannot run power hungry GPS radios on the tiny batteries that small insects such as bees can carry. 
 \item Accurate wireless tracking relies on phase information~\cite{phaser,dinansdi,spotfi} which requires a radio receiver at the insect.  This is challenging  because radios are power consuming. Further, we are unaware of  small low-power off-the-shelf radios that provide I/Q samples of the raw RF signals or CSI data.
\item Existing localization algorithms are designed for  Wi-Fi chips~\cite{dinansdi,phaser} or software radios that are not constrained by their computational capabilities~\cite{robotkyle,dina2014,sensys2018}. In contrast, self-localization at the insect requires algorithms that can operate with a tiny antenna and a low-power microcontroller.
\squishend

Our design instead eliminates power-hungry radio receiver components (e.g., high frequency oscillators) by using passive operations to perform envelope detection and  extract  just  the  signal  amplitude. While this design enables the insect to receive at a low-power, it discards the phase information, which is essential for  RF localization.

{We present a novel technique that {\it extracts the angle to the AP from the amplitude information output by the envelope detector.}} Our intuition is as follows: say the AP broadcasts signals to  the insects from two of its antennas. By changing the relative phase between the two transmit antennas at the AP, we can create amplitude changes at the insect's envelope detector over time. These amplitude changes effectively give us multiple equations that allow us to solve for the angle to each AP. Combining the angle information from two APs allows our platform to localize itself on a farm using a passive envelope detector. In~\xref{sec:loc}, we build on the above intuition and present a low-complexity  algorithm that works in the presence of multipath as well as at speeds of up to 9.1~m/s.

{\bf Summary of results.} Fig.~\ref{fig:bee1} shows our hardware including the antenna, backscatter communication, wireless receiver for self-localization and the circuit board that connects the microcontroller and sensors in a lightweight form factor using micro-fabrication techniques (see~\xref{sec:fabrication}). We attach our 102~mg platform, including a 70~mg battery, to three common species of bumblebees (\textit{Bombus impatiens, Bombus vosnesenski, and Bombus sitkensis}). 

Our results show the following.
\squishlist
\item {\it Communication.}  Our microcontroller-based backscatter design, placed on the bee, can transmit bits using ON-OFF keying at 1~kbps, when the bee returns to the hive. 
\item {\it Self-localization.} Across deployments in a soccer field and farm, our envelope detector design on the bees achieves a median angular resolutions of $4.6^{\circ}$ at ranges up to 80~m. Ground-truth benchmarking with drones show that similar angular resolution can be achieved even at speeds of 9.1~m/s. 
\item {\it Power.} With the on-board 70~mg rechargeable battery, our system could last up to 7~hours while sampling its location once every 4 seconds. We also show the feasibility of  fully recharging the battery back at the hive within 6~hours, using RF power. 
\item {\it Sensing.} We build prototypes including humidity, temperature, and light intensity sensors. These sensors fit within our 102~mg prototype, enabling mobile sensing using bees.
\squishend

{\bf Contributions.}  While we are not the first to place electronics on insects --- prior work in biology uses electronics on bees to understand their foraging behavior~\cite{surveyinsects,beeart1,beeart2} --- {we show for the first time that insects such as bees can be used to carry general purpose sensors,} localize themselves and perform location-based sensing operations.

To summarize, we make the following conceptual, technical and systems contributions. First, we explore the idea that insects can be used in lieu of drones to enable mobility for sensor networks. Second, we present a novel general-purpose  platform that is low-weight and can support computing, communication and sensing operations on flying insects. Third, we introduce  the first  self-localization technique for small insects like bees using a novel algorithm that computes 2D location using  only the output of a passive envelope detector.


\section{Application Scenarios}

Smart farming techniques utilize a variety of sensors to measure plant health. For example, moisture and humidity sensors measure the availability of water, light sensors measure the availability and intensity of sunlight, and temperature sensors can help determine whether conditions are optimal for particular crops. Live insects like bees present an attractive option for use as mobile sensing platforms for agriculture.  Bees are among nature's best pollinators and are regularly purchased for use on farms~\cite{koppert}, and many fruit crops \textit{depend} on bees for pollination. These insects fly for hours foraging for food, and also fly up to individual plants, which is difficult to do with drones. Further, tracking bees alone could give important insights about pollination which is not available from commercial sensors. This includes pollination patterns that can help maintain genetic diversity~\cite{russell1, russell2}.  

\begin{figure}[t]
\includegraphics[width=0.9\linewidth]{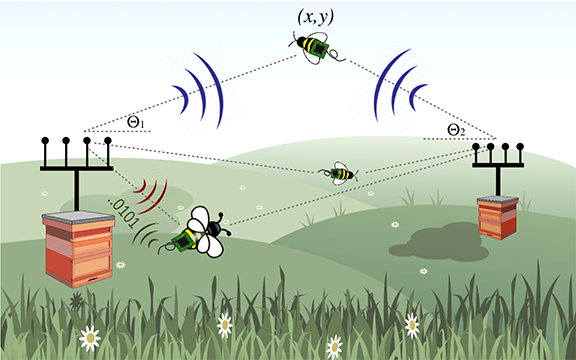}
\caption{\small{Insect-borne sensor packages can self-locate and collect location-dependent data using on-board sensors. The data are uploaded to AP  when the bee is back at the hive. }}
\label{fig:bee_illustration}
\end{figure}

We outline two deployment scenarios: 1) the sensors periodically wake up from a low power sleep mode to sample sensor data and location, store the data in memory, and then return to sleep mode. As the insect goes about its business foraging for food, we get samples of data along its trajectory. This is ideal for low power sensors such as temperature, humidity, etc. that simply require storing a single value, and 2) we can imagine other applications where we wish to acquire more detailed information at a particular location that is programmed when the bees are the hive. In this case, the system could periodically check its location, and \textit{opportunistically} sample the sensor if it is close to the destination. 


\section{\Name\ Design}
The idea of using flying insects as a mobile sensing platform raises a number of wireless networking and sensing challenges. At a high level, this requires an ultra-miniature sensing and computation package, memory to store data, a power source, and wireless connectivity for data transfer. Additionally, it requires a wireless localization method capable of connecting the sensor measurements to specific locations.

Our solution consists of two components: a lightweight battery powered electronics package including a sensor, RF switch used for backscatter communication and microcontroller that can be mounted on a flying bee, and multi-antenna access points with dedicated power sources that broadcast the RF signals needed for backscatter and localization.
 Since large bumblebee hives cover foraging areas of roughly $8,000$ $m^2$ \cite{greenmethods}, we target an operating range of 50--100~m for  localization. The range requirements for communication however are much smaller as sensor data can be stored in onboard memory and uploaded when the bee returns to the hive.

To achieve this, our bee-mounted electronics package requires the following two components: 
\squishlist

\item  {\it Low-power self-localization.} The first requirement is a localization algorithm that runs  on the insect mounted platform and computes its location based on wireless transmissions from APs at known locations. By adopting this broadcast architecture similar to GPS, any number of insects can concurrently compute their location.  We present a novel algorithm that uses a passive envelope detector to receive the RF signal and compute the location on a microcontroller.

\item  {\it Communication.} The second requirement  is a form factor compatible wireless communication link to download sensor data. Our backscatter approach has the advantage of requiring only an antenna connected to RF switches and a microcontroller, all of which can be achieved using off-the-shelf components. 

\squishend

In the rest of this section, we first describe our insect form factor compatible platform. We then present algorithms to estimate the  location from the output of the envelope detector. Finally, we describe our backscatter design for uploading data back to the AP when the bee returns to the hive.

\subsection{Form-Factor \Name\ Platform}
\begin{figure}[t!]
\vskip -0.16in
\includegraphics[width=0.4\linewidth]{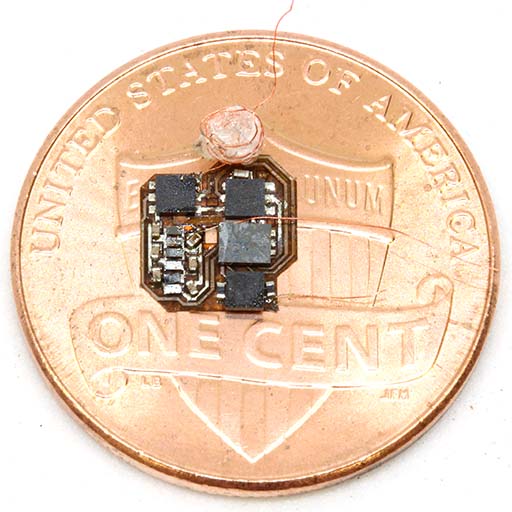}
\hspace{.2in}
\includegraphics[width=0.23\linewidth]{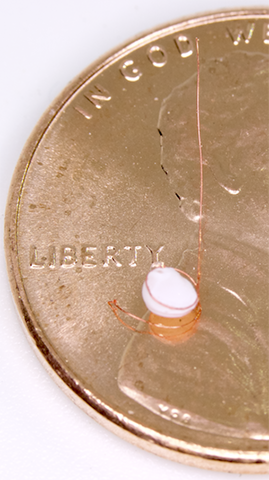}
\caption{\small{Complete electronics package including an antenna, envelope detector, backscatter transmitter, and sensor shown on a US penny for scale (left). Custom miniature antenna (right).}}
\label{fig:antenna}
\vskip -0.06in
\end{figure}
\subsubsection{Understanding form-factor requirements.}
We choose bumblebees as it has been well documented that they can fly while carrying payloads of their own body weight or more. We purchased a commercially available \textit{Bombus impatiens} colony~\cite{koppert}. We progressively added weight to the insects and observe that healthy workers measuring roughly 13~mm in length are capable of controlled flight while carrying loads of approximately 105~mg. When adding weight beyond this limit, they are unable to perform controlled hovering and have difficulty taking off. We perform similar experiments with wild bees specifically \textit{Bombus vosnesenski} (Yellow faced Bumble bee) and \textit{Bombus sitkensis} (Sitka  Bumble bee). We note that these insects are slightly larger, 14 and 16~mm respectively and can carry slightly more weight. With the weights noted above, the insects were active and exhibited normal behavior.
Thus, we target 105~mg or less for our platform. Of this, 70~mg is consumed by our  3V 1~mAh rechargable lithium ion battery~\cite{battery80mg}. This allows only 35~mg for communication, computing, sensing and self-localization.

\subsubsection{Fabrication Method}\label{sec:fabrication} Our platform consists of four different elements: a microcontroller, RF switches, an envelope detector, and sensor. For each of these we leverage commercially available components available in ultra-miniature packages. The core of our design is the Kinetis KL03z ARM Cortex M0+ microcontroller~\cite{micro1} which is available in an 2 x 1.61~mm package and weighs only 4.1 mg. We use this microcontroller to sample the output of the envelope detector and the sensor, and to toggle the RF switches for backscatter communication. We use two Skyworks 13314-374LF single pole dual throw switches weighing 3.3~mg each, the first to select between the envelope detector and backscatter, and the latter to toggle between the two backscatter impedances. We construct the envelope detector out of small diodes and capacitors consuming a total area of 7.26~$mm^2$. We test different sensors including a TI HDC2010 humidity and temperature sensor as well as an ALS PT19 photodiode to measure light intensity. In total our whole platform weighs 102~mg and measures $6.1\times 6.4~mm^2$. 

We must also consider the weight of the substrate such as the printed circuit board (PCB) used to create the circuit that connects these components. The weight of a $8\times 6\times 3.175$~mm PCB made of copper clad FR4 with standard thickness and density of 2.6~g/cm$^3$ is greater than 390~mg. We instead fabricate our own light-weight PCBs by laser micromachining 0.5~oz copper coated sheet of 127~um thick FR4 and the result is illustrated in Fig.~\ref{fig:antenna}. We begin by cleaning the copper with isopropanol and placing it on a low tack adhesive~\cite{gelpak}. We then raster out the desired pattern of copper traces using our laser micromachining system~\cite{dpsslaser} followed by a low power cleaning raster over the resulting copper traces to remove dust and cut away the excess material. The resulting circuit is approximately 100~$\mu$m thick. We then populate the board with components and solder them.


\begin{figure}[t!]
    	\centering
\includegraphics[width=0.35\linewidth]{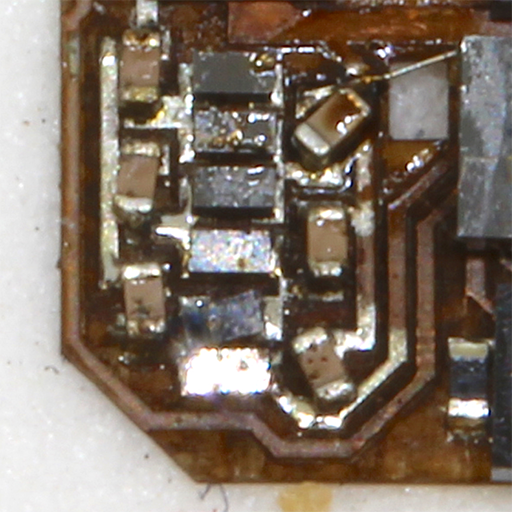}    	
\includegraphics[width=0.55\linewidth]{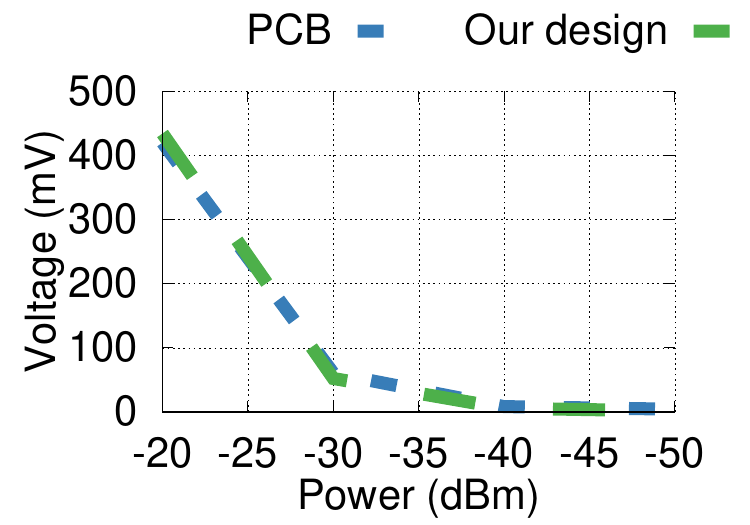}
  \vskip -1em
  \caption{\small{\textbf{Envelope detector performance,}}  {comparing a standard thickness PCB to our light-weight design (left) showing it does not compromise performance.}}
  \label{fig:env_benchmark}
  \vskip -0.05in
\end{figure}

\subsubsection{Envelope detector}
\begin{figure}[t!]
    	\centering
    	\includegraphics[height=2.5cm]{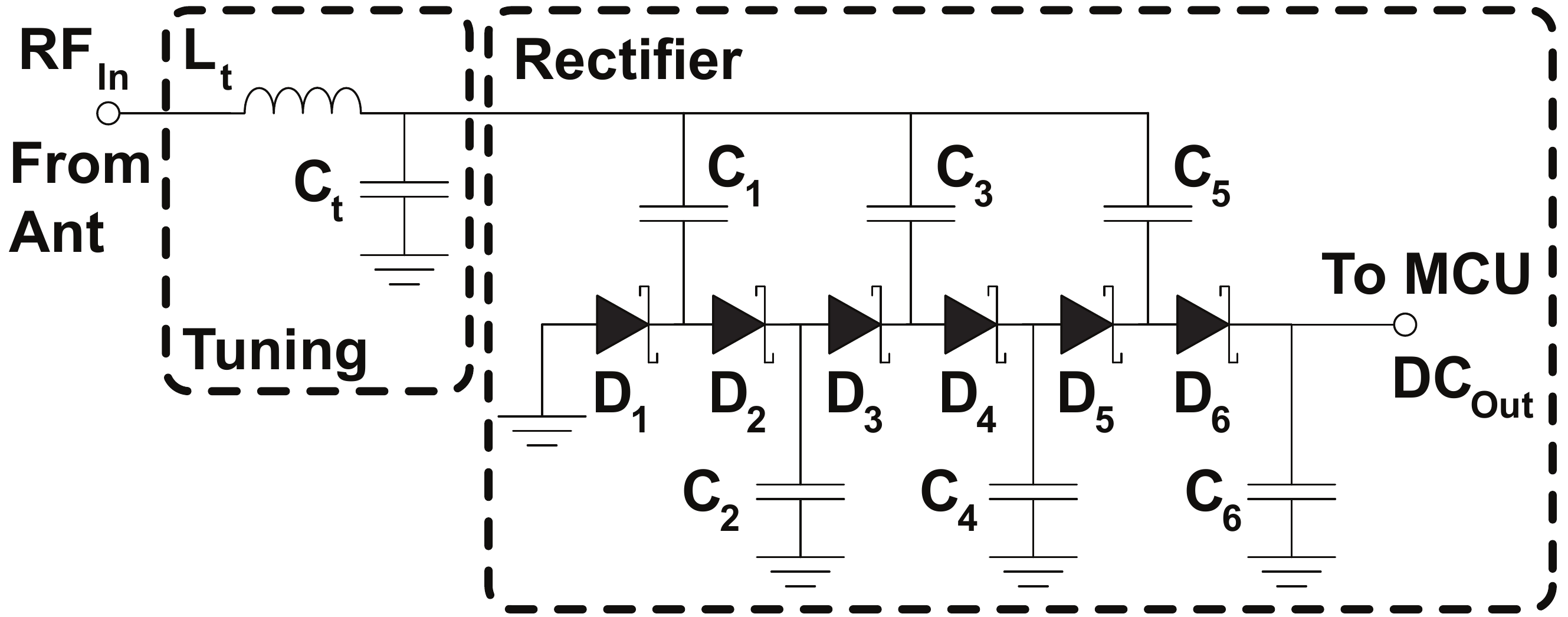}
  \vskip -1em
  \caption{{\small{Envelope detector circuit diagram.}}}
  \label{fig:env_circuit}
  \vskip -0.05in
\end{figure}

{To design our envelope detector circuit, we use the 3 stage rectifier design shown in Fig~\ref{fig:env_circuit} that is presented in~\cite{multiband-harvester}.} In order to minimize size and weight, we instead use small SMS7630-061 zero bias silicon Schottky detector diodes which are available in a 0.3x0.6~mm package and 0201 capacitors each weighing less than 0.5~mg~\cite{diodes} and tune the circuit for operation in the 915~MHz ISM band. To understand whether our miniaturized circuit compromises performance, we compare its performance to the one in~\cite{multiband-harvester} made with larger components on standard thickness PCBs. We use a USRP  to transmit a tone at 915~MHz and measure the rectifier output voltage with a digital multimeter. Fig.~\ref{fig:env_benchmark} shows the rectifier output voltage versus input power for the two designs, demonstrating we achieve comparable performance. {Our receiver  sensitivity of -40~dBm is similar to the values achieved in~\cite{lorabackscatter}.}

\subsubsection{Antenna design}
For simple antenna geometries, the frequency of operation directly affects the size of the required antenna. Typical antennas are a half or quarter of the wavelength in size. At 915~MHz, the wavelength is approximately 33~cm, making these antennas significantly longer than a bumblebee (8.5-16~mm)~\cite{beesize}. We explore the design space to achieve a high performance small and light-weight antenna. 


First we consider chip antennas that attempt to miniaturize antennas by using materials such as high dielectric constant ceramics. While a 900~MHz chip antenna may only be $3.2\times 10~mm$, they often require a large ground plane to perform efficiently. For example, the recommended ground plane for the above chip antenna is $50\times 120$~mm, and the antenna requires clearance to other components~\cite{ant900}. Additionally, the dense ceramic materials increase the weight making the antenna 288~mg.  An alternative is a wire antenna which is a significant fraction of a wavelength. Although these may be longer than the insect~\cite{beeradio}, the wires could hang off the sides or back. A 15~cm long straight wire hanging off the sides however could introduce other practical issues in the context of insect flight --- thin wires can be easily tangled and caught in the insect's wings, legs, or even plants since they don't retain a rigid shape.

To address these concerns we instead design a base loaded whip antenna constructed using the 43~AWG wire. This antenna design is a common solution used in cars and handheld transmitters with electrically short antennas. Rather than having just a straight wire whip which with shortened length presents capacitive reactance, adding inductance at the base cancels this and creates a resonant antenna. We fabricate this antenna using 7 turns of 43~AWG wire with a diameter of 2~mm followed by a straight 10~mm  wire. We then apply cyanoacrylate glue (Loctite 416) to the antenna to stiffen it and prevent it from losing its structure. {The resulting structure is shown in Fig.~\ref{fig:antenna}, has a total weight of 4.6~mg, and length of 10~mm which is smaller than the bee and avoids the above issues of entanglement. We compare this to the performance of a 900~MHz monopole antenna. We place each antenna 3~m from a transmitter and find that our lightweight design receives roughly 6-8~dB lower power.}

\subsubsection{Attachment to Live Insects}
We begin by capturing individual bees in plastic jars. For our \textit{B. impatiens} colony we open the door to the hive and place a small cotton swab dipped in sugar syrup near the hive exit to lure bees out one at a time and trap each insect in a jar. For the other species, we capture live bees from a small farm, again by trapping them in a plastic jar while they forage on flowers. After isolating a single insect, we place the plastic jar in a freezer at approximately 0$^{\circ}$C for 4-5~min to cold anesthetize the insect~\cite{pollinationbook}. At this point the bees typically stop moving allowing for 2--5 min of working time to attach the electronics. We then grip the insect by the sides of the thorax with a pair of tweezers and adhere our electronics package by applying a drop of cyanoacrylate glue and quickly applying a small amount of an accelerator compound (Loctite SF712) to immediately dry the glue and hold it in place. We experiment with attaching the electronics in two locations: the top of the thorax behind the head and on the upper abdomen. We find that the bees are able to fly when the weight is attached in either position, however we prefer the abdomen as it naturally droops down away from the wings when the bee is held off the ground and minimizes the risk of excess glue flowing onto the wing joints on the thorax.


\subsection{Self-localization of insects}
\label{sec:loc}

Living IoT's localization system employs APs which transmit RF signals in the 900~MHz ISM band. At the insect-mounted receiver, due to the power and size constraints, \name\ uses a passive envelope detector connected to a small antenna to receive only the amplitude of the RF signals broadcast by the APs.  Unlike an active radio that gives both amplitude and phase, the envelope detector only provides the amplitude of the signal but can do so with small, zero-power passive hardware components~\cite{allsee,barnet}.  The phase of the signal is however essential to achieve wireless localization. 

To address this, we first extract the phase difference of the signals from each of the transmit antennas of an AP using the amplitude output by the envelope detector. We then use this phase to compute the angle of the insect with respect to that AP. By using the angles from two APs we can identify the 2D location. Typically bees only fly  a few meters above the ground and hence 3D localization is not essential. However, the technique presented in this paper can be generalized to achieve 3D localization by adding an additional AP. Next, we describe each of these techniques in further detail.

\begin{figure}
\includegraphics[width=0.7\linewidth]{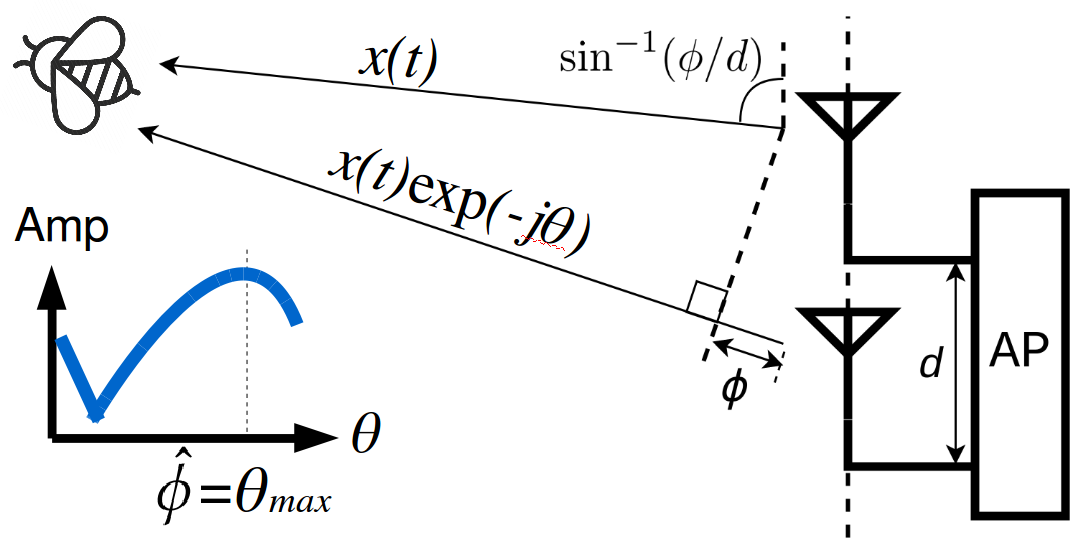}
\caption{\small{Self-localization using two antennas at AP.}}
\label{fig:bee_localization}
\vskip -0.05in
\end{figure}
\subsubsection{Phase from Envelope Detector}
\label{subsubsec:phase}
Consider a setup in which an AP transmits RF signals using two of its antennas as shown in Fig.~\ref{fig:bee_localization}. The two transmissions will travel different distances to the receiver and therefore combine at the insect with a phase difference corresponding to this distance difference. The small antenna at the insect receives the combined RF signals and the envelope detector outputs its amplitude. The key insight here is that the  amplitude  depends upon the phase difference at which the two transmitted signals combine at the receiver. For example, if the transmitted signals are perfectly in-phase they will add constructively giving the maximum amplitude; in contrast a phase difference of $\pi$ will cause the two signals to cancel each other completely.

Our key insight is that by intentionally introducing an additional phase difference between the two AP antennas, we can create amplitude changes at the receiver. We can analyze these  changes to estimate the phase corresponding to the angle of the insect from the AP.

To explain this in more detail, 
consider $x_1(t)$ and $x_2(t)$ to be the signals transmitted from the two antennas. Let us set both these signals to  $x(t)=Ae^{j\omega t}$. Assuming no multipath which we will discuss later, the signal at the receiver envelope detector can now be written as:
\begin{equation*}
y(t) = |ax_1(t) + ae^{j\phi}x_2(t)| 
=aA(\sqrt{2+2\cos(\phi)}) 
\end{equation*}

Here $a$ is the signal attenuation, and $\phi$ is the phase difference between the two paths. Note that the amplitude attenuation difference between the two signals is negligible since the separation between the two antennas is small compared to the distance between the AP and bee.

Now if the AP intentionally introduces a phase difference of $\theta$ on the second transmit antenna, i.e., $x_2(t)=x(t)e^{-j\theta}$, the signal at the receiver can be written as,
\begin{equation*}
y(t) = |ax(t) + ae^{j(\phi-\theta)}x(t)| = aA(\sqrt{2+2\cos(\phi-\theta)}) 
\end{equation*}

The maximum value for the above equation happens when $\phi-\theta=0\ mod\ 2\pi$. Hence, to get an estimation $\hat{\phi}$ of $\phi$, we can let the AP sweep $\theta$ from $-\pi$ to $\pi$ at a constant rate. At the receiver side, the envelope detector simply samples the amplitudes corresponding to each of the $\theta$s. It then gets the sample with the maximum amplitude and infers $\theta_{max}$ from the time of that sample. Now, the $\hat{\phi}$ we are interested in is simply $\theta_{max}$.
The angle $\hat{\Theta}$ of arrival of the receiver corresponding to the two transmit antennas at the AP can be derived by $\hat{\phi}=d\sin(\hat{\Theta})$ where $d$ is the distance between the two transmit antennas in radians. In our design, this distance between transmit antennas is set to half a wavelength.

A key consideration in the above design is that as the distance increases, noise affects the signal quality. We mitigate the effect of this noise by sampling for a longer time, i.e., the AP dwells on each phase difference $\theta$ for a longer duration. The upper bound on this duration however is determined by the motion of the bee. Our empirical results found that
a duration of 50~ms per sweep across all the angles, is a good trade-off between the noise level and motion tolerance.

\subsubsection{Addressing Multi-path.} The above discussion assumes that the phase difference $\hat{\phi}$ estimated at the insect can be translated into angle of arrival using amplitude variations.  We however need to address the potential amplitude variations due to multipath.  Unlike systems like Wi-Fi which operate indoors with mostly no LOS path, our system is designed for outdoor farm use in the natural habitat of insects. For example, in deployment scenarios such as open fields and farms there is a direct strong line of sight. However, we still need to account for the amplitude variations that result from the other paths constructively and destructively interfering with the dominant direct line of sight path. To this end, we utilize more than two antennas per AP to reduce the error in the angle caused by multipath.

Formally, suppose we use N antennas separated by half a wavelength each. Similar to the above two-antenna scenario, we introduce a phase offset  of $\theta_i$ for the $i$th antenna. Suppose $M$ independent multipaths exist and the first one is the direct path, the
received signal can be represented as, $y_{env}(t) = |A\sum_{k=1}^M a_k \sum_{i=0}^{N-1} x(t)e^{j(i*\phi_k-\theta_i)}|$. Here $a_k$ is the attenuation for the $k$th path. Let us set  $\theta_i=\theta*i$, and sweep $\theta$ from $-\pi$ to $\pi$. We now get, $y_{env}(t) = A|\sum_{k=1}^M a_k \sum_{i=0}^{N-1} e^{ji(\phi_k-\theta)}|$.

We solve for $\theta_{max}$ using the same procedure as in the 2-antenna case and estimate $\hat{\phi}=\theta_{max}$, where the amplitude output by the envelope detector has the maximum value.

\begin{figure}[t!]
\includegraphics[width=0.9\linewidth]{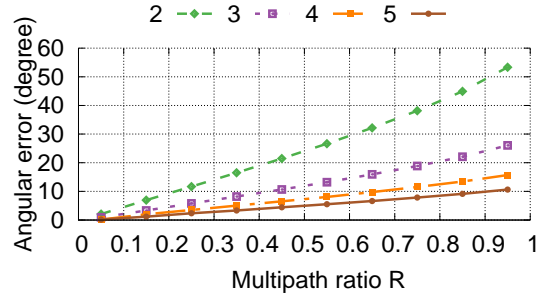}
\vskip -0.1in
\caption{\small{Simulation results for different multipath ratios,  $R=\frac{\sum_{i=2}^M a_j}{a_1}$,  using 2--5 transmit antennas.}}
\vskip -0.15in
\label{fig:multipath_sim}
\end{figure}

The key insight is that, when $\sum_{j=2}^M {a_j}<a_1$, which is true in most line-of-sight scenarios like the farm, the error in our estimate of the angle with respect to the AP, which is $|\hat{\phi}-\phi_1|$, is bounded. Moreover, this error decreases linearly as the number of antennas increases. To verify this intuition, we perform a simulation where we compute this error by changing the number of antennas. {We repeat this for increasing multipath ratios $R$, ratio of the sum of the amplitudes of all NLOS paths with respect to the amplitude of the LOS path.} Assuming that the angles of indirect paths are uniformly distributed, Fig.~\ref{fig:multipath_sim} shows the mean error as a function of these two parameters. The plot shows that the error is less than 10$^\circ$ when using four antennas even if the total amplitude of all NLOS path is 60\% of the amplitude of LOS path. With five antennas, we can get a similar error even when this ratio is close to 0.95.  {This shows that by increasing the number of antennas at the AP, we can reduce the error due to NLOS paths and achieve an accurate angle estimation.}

\subsubsection{Leveraging insect motion.}
We also leverage insect motion to reduce the effects of multipath and reduce the probability of small scale fading. This is specifically useful when the bee is in motion. Since the typical speeds when the bee is in motion are less than 10~m/s, the bee does not move by more than a meter between consecutive 50~ms durations where our AP cycles across the phase values.  Despite introducing slight errors because of Doppler effects, our algorithm utilizes motion to improve the accuracy by leveraging spatial diversity: the multipath combination can be significantly different for even a small displacement. Hence, we use exponential smoothing to temporally average consecutive measurements which yields a more accurate result. Formally, the final angle of arrival $\Theta_t$ of each bee is calculated as: 
$\Theta_t=\eta \Theta_{t-1}+(1-\eta)\sin^{-1}(\hat{\phi}_t/\pi)$, where $\eta$ is the smoothing constant which we set to 0.8. Note that the computationally expensive $\sin^{-1}$ operation can be offloaded to the access point, as shown in Algorithm~\ref{fig:alg_ap}.

\begin{algorithm}[t]
    \caption{Pseudo-code at the  AP.}
	\SetKwInOut{Input}{Input}
	\SetKwInOut{Output}{Output}
	\For{$AP_i$ in $\{AP_1, AP_2\}$}{
    $AP_i$ transmit $preamble_i$ \\
    \For{$\Theta=-\pi/2$ to $\pi/2$ by $\delta$}{
    \For{$j=2$ to $4$}{
    set the phase shift of the $j$th phase shifter to $(j-1)*\pi*sin(\Theta)$ \\
    }
    sleep for $(T-T_{preamble})/\pi*\delta$ \\
    }
    }
    \label{fig:alg_ap}
\end{algorithm}

\subsubsection{2D localization of insects}
To estimate the 2D location of the insect, we employ two APs with four antennas each. Separating the two APs and placing them perpendicular to each other gives the best 2D accuracy. Given the known locations of the two APs, the two angles computed with respect to each AP, uniquely identifies the 2D location. The bees can either store these two angles or can also calculate their 2D location of using the intersection of the two separate angles of arrival. This intersection procedure can be implemented using a look-up table to minimize the required computation.
 
Specifically, the two APs intermittently transmit their sweep signal one after another. {They are coarsely synchronized using TDMA so that no two sweeps will interfere with each other. We transmit two predefined orthogonal preambles using ON-OFF keying, [1,0,1,0,1,0,1,0] and [1,1,0,0,1,1,0,0] to identify each AP.} The preambles are transmitted before every sweep of each AP respectively so that the bee can find the start of each sweep efficiently. The receiver first detects the preamble using a simple state machine, then runs the algorithm twice to get the two angles. Our pseudo-code  is presented in  Algorithm~\ref{fig:alg_rec}. We note that the computation for our receiver algorithm scales linearly with the sampling rate (which is in the order of kHz), and only simple arithmetic operations are involved. This makes it efficient enough to run on our microcontroller platform. 

A key consideration while increasing the number of APs is that the transmissions from each of them have to be time-multiplexed. This increases the delay required to compute the location value which can be challenging especially when the bee is highly mobile. Given that each AP sweeps across various phase values in 50~ms segments, the delay to compute the 2D location  is around 100~ms. Assuming a speed of 3~m/s, this translates to a motion of around 30~cm which is within the error of our location estimates and hence does not significantly affect our accuracies. 

\begin{algorithm}[t!]
	\SetKwInOut{Input}{Input}
	\SetKwInOut{Output}{Output}
	\While{True}{
    receive and store amplitude samples during the last $3T$ time into memory as $S$ \\
    \If{$preamble_1$ detected in $S_{1..2T}$ located at $i$}{
    $p_1=argmax_{j\in (T_{preamble}, T)}S_{i+j}$\\
    $angle_1=(p_1-T_{preamble})/(T-T_{preamble})*\pi-\pi/2$\\
    locate $preamble_2$ in $S_{i+T..3T}$ at $i'$\\
    $p_2=argmax_{j\in (T_{preamble}, T)}S_{i'+j}$\\
    $angle_2=(p_2-T_{preamble})/(T-T_{preamble})*\pi-\pi/2$\\
    (If Euclidean position is needed) get $x$ and $y$ from $angle_1$ and $angle_2$ using look-up table \\
    output ($angle_1, angle_2$) or ($x$, $y$)\\
    }
    }    
    \caption{Pseudo-code of Living IoT platform.}
    \label{fig:alg_rec}
\end{algorithm}

\subsubsection{Localization power consumption.} 
In order to minimize power consumption, we duty cycle our localization and sensor by periodically performing a measurement and then returning to a low power mode until the next measurement. Sampling the envelope detector and performing computation for localization requires a peak current of 1.6~mA followed by return to a low power sleep mode with a current of 100~$\mu$A. If we check our 2D location every 4s, the average power consumption is  138~$\mu$A. For our battery capacity of 1~mAh, this results in potential battery life of up to 7~hrs on our rechargeable battery. 

\subsection{Backscatter Communication Design}

\begin{figure}[t!]
\begin{subfigure}{0.5\columnwidth}
    	\centering
    	\includegraphics[width=1\linewidth, height=2.5cm]{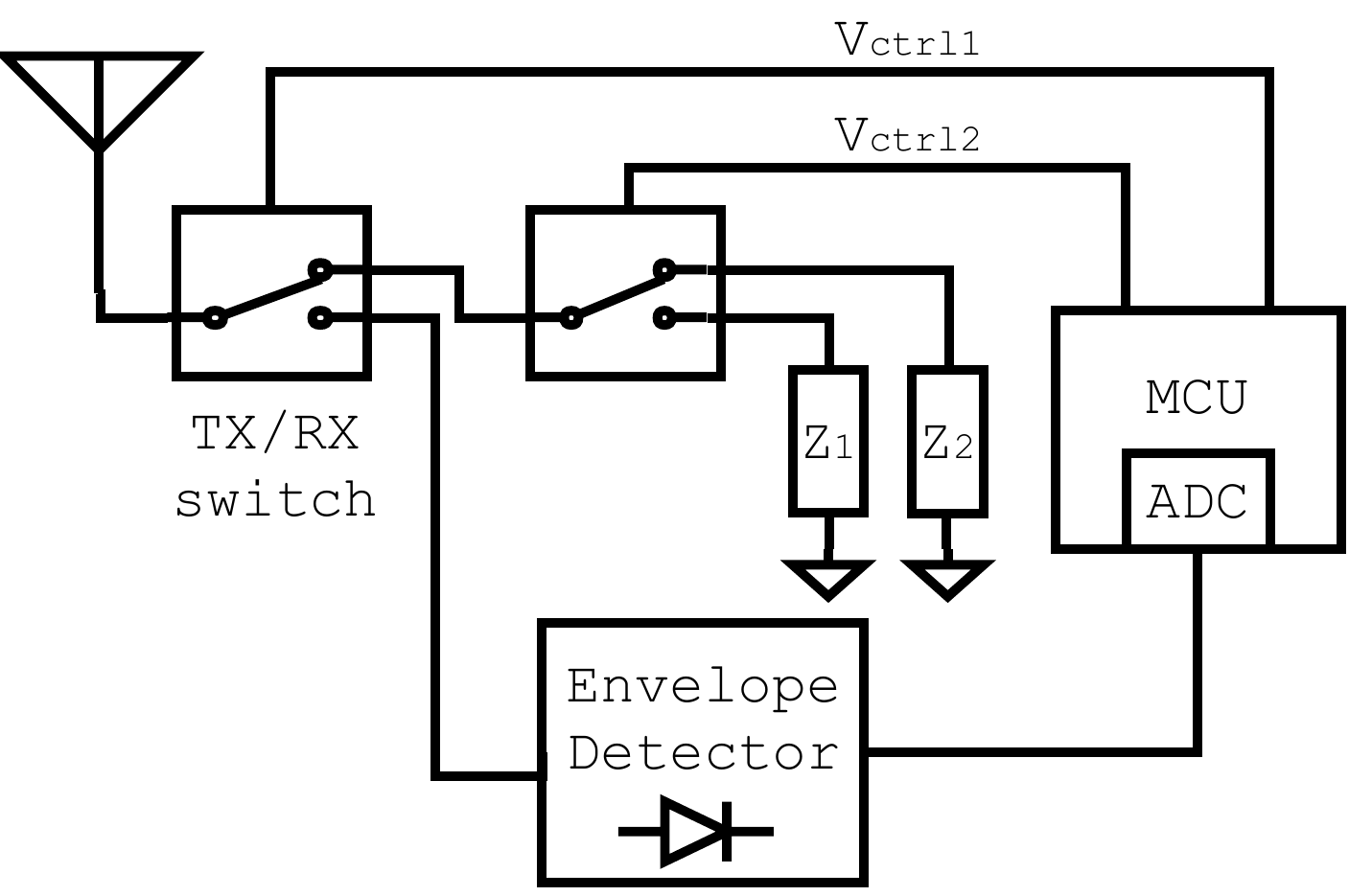}
	\end{subfigure}
    \begin{subfigure}{0.4\columnwidth}
    	\centering
    	\includegraphics[width=1\linewidth]{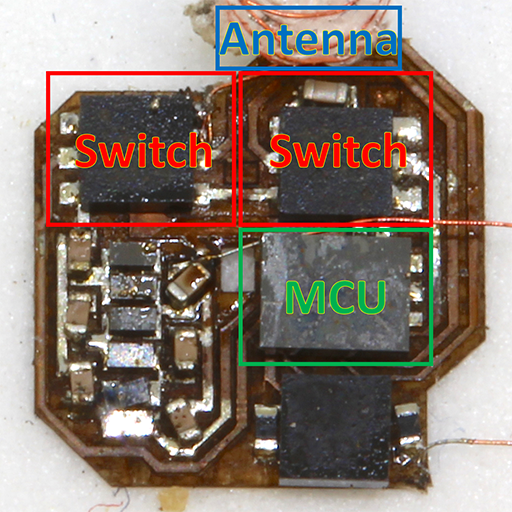}
	\end{subfigure}
  \vskip -1em
  \caption{\small{Backscatter hardware including a block diagram (left) and light-weight hardware implementation (right).}}
  \label{fig:backscatter_design}
  \vskip -0.05in
\end{figure}

Backscatter requires three components: an antenna, a switch to modulate the reflected signal, and a control signal for the switch with the  desired bits~\cite{interscatter,abc}, two RF switches, and the microcontroller as shown in Fig.~\ref{fig:backscatter_design}. We connect the antenna to the input port of the first switch which selects between the envelope detector and the second switch used for backscatter. We use the second switch to select between two impedance states to create the backscatter signal. In order to minimize extra components we simply use the open and short impedance states which we implement by leaving one port disconnected and connecting a 51~pF capacitor to ground on the other port. 

At a high level, the AP transmits a continuous wave signal from one of its antennas, which the \name\  platform backscatters with the sensor data when the bee is back near the hive. We use ON-OFF keying modulation to encode bits using backscatter.  Backscatter however presents the challenge of needing to receive the weak reflected signal in the presence of the high power RF signal from the AP. To address this using a technique called subcarrier modulation~\cite{interscatter,nsdi16}: we use the microcontroller to generate a square wave at a rate of 2~MHz which shifts the frequency of our backscatter signal 2~MHz away from the high power signal from the AP. By doing this, the AP can then  filter out the transmitted signal thereby significantly reducing the receiver noise floor and improving our communication range. We then modulate this signal by toggling it ON and OFF to produce an ON-OFF keying pattern to encode bits.

To implement this with minimal power consumption, we disable all unnecessary peripherals on the device and reduce the clock frequency to 8~MHz. We then generate a 2~MHz square wave using a timer module in PWM mode at a fixed 50\% duty cycle for sub-carrier modulation. We find that enabling and disabling the timer module in software incurs delays, so we instead take advantage of our first switch: we use a lower rate control signal from a GPIO pin on the microcontroller to toggle the first switch to the envelope detector and disconnect the backscatter signal. {Using this method we can send data at a rate of 1~kbps which is sufficient for our application where the sensor readings and angle values we need to transmit are only 1-2~bytes.}

\begin{figure}[t!]
\begin{subfigure}{0.35\columnwidth}
    	\centering
    	\includegraphics[width=1\linewidth,height=2.8cm]{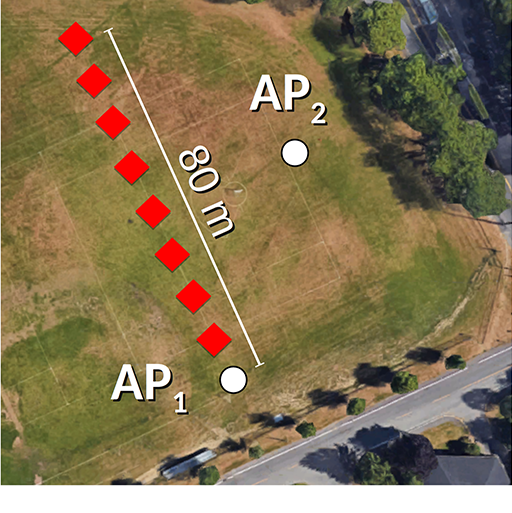}
	\end{subfigure}
    \begin{subfigure}{0.6\columnwidth}
    	\centering
    	\includegraphics[width=1.05\linewidth]{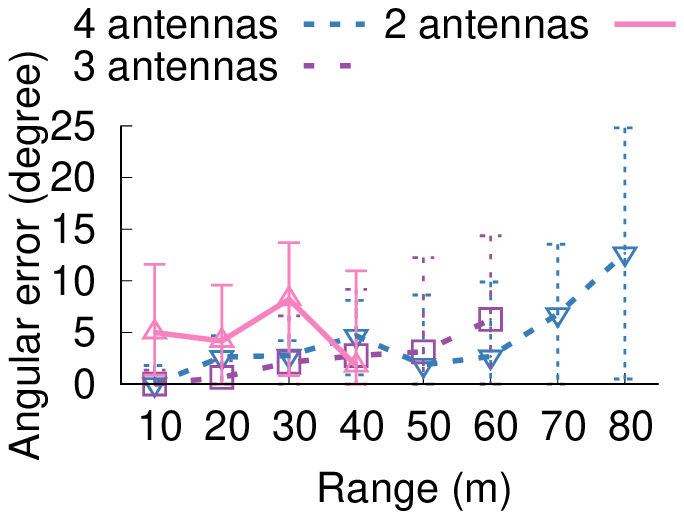}
	\end{subfigure}
  \vskip -1em
  \caption{\small{{Stationary self-localization angular accuracy.}}}
  \vskip -0.05in
  \label{fig:1drange}
\end{figure}

We note the following about our communication design.
\squishlist
\item {\bf Backscatter range.} In our design the insects store the sensing data and upload it to the AP via backscatter when the insect is back at the hive. This only requires a range of a few meters. One can, in future designs, consider increasing the backscatter range using coding to hundreds of meters~\cite{lorabackscatter}. 
 
\item {\bf Downlink and MAC protocol.} We reuse the envelope detector  for downlink communication from the AP to the bees. Our design uses a simple MAC protocol where, once the bees are back at the hive in the night, the AP queries each of the insects one after the other, using the downlink. The insect then uses backscatter to respond with the sensor data and corresponding location information. 
\item {\bf Effect of interference.} A concern with using an envelope detector is that it is not frequency selective and can therefore see signals across a broad range of frequencies at 900~MHz. This however does not significantly affect our performance for two key reasons. First, our application is designed for smart farms where 900~MHz transmitters including cordless phones, LoRa and RFID readers are currently uncommon. Second, most 900~MHz wireless deployments including LoRa and SIGFOX  are designed for sensitivities of less than -118~dBm. These lower power signals do not register at our envelope detector which has a -40~dBm  sensitivity. 
\item {\bf Power requirements.} 
Our micro-controller based backsc-     
atter requires a peak current of 1.8~mA when transmitting at 1~kbps. Unlike localization which is performed periodically, data is only uploaded once upon returning to the hive and does not otherwise consume power. {Offloading 10 sensor measurements and the corresponding angle data would only require running our backscatter transmitter for 32~ms} and is therefore not a concern from a power perspective.
\squishend
                                                                                                                                                                 
\section{Evaluation}
We evaluate various aspects of our \name\ platform.
\subsection{Self-Localization Accuracy}


\begin{figure}[t!]
\begin{subfigure}{0.5\columnwidth}
    	\centering
\includegraphics[width=1\linewidth]{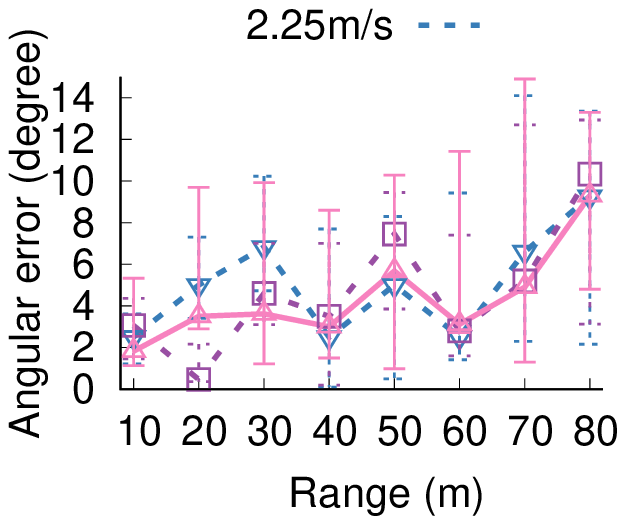}
        \caption{AP1}
	\end{subfigure}
    \begin{subfigure}{0.45\columnwidth}
    	\centering
    	\includegraphics[width=1\linewidth]{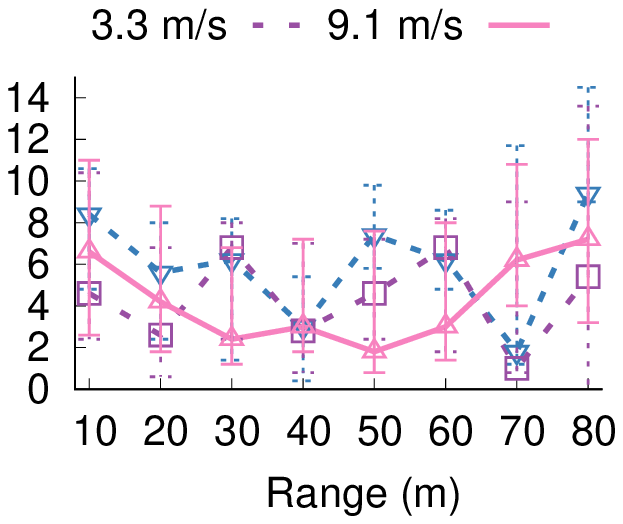}
       \caption{AP2}
	\end{subfigure}
  \vskip -1em
  \caption{\small{{Mobile 2D localization angular accuracy.}}}
  \vskip -0.05in
  \label{fig:drone}
\end{figure}


We first evaluate our low-power localization algorithm using a static deployment of \name\ in different environments. Since it is difficult to get the ground truth location with a free flying bee, we use a drone to evaluate the 3D accuracy of our technique at different speeds. Then, we place the living IoT platform on a live, wild caught bee that can fly freely within a plastic enclosure and run our localization algorithm at various points across a farm. Finally, to benchmark the effect of different wing speeds, we run experiments with a robotic wing of similar dimensions. 

\subsubsection{Stationary experiments.} \label{subsubsec:2dlocation} For our experiments, each AP consists of one USRP-N210 connected to a four way power splitter followed by three phase shifters, each of which introduces a phase shift controlled by an NI myDAQ digital-analog converter. Along with the original signal, the four outputs are amplified {to 28~dBm} by a Qorvo RF5110G power amplifier and then connected to {four 2~dBi monopole antennas} separated by 12~cm each. The cable lengths are carefully calibrated so that no extra phase offset is introduced.


{\bf Soccer field deployment.} We first conduct our range evaluation outdoors on an open soccer field measuring approximately $100\times 100$~m. We place an AP at one end of the field and move our receiver along a straight line away from the AP up to a maximum distance of 80~m. {Fig.~\ref{fig:1drange} plots the angular error as a function of distance up to the point at which the receiver had insufficient SNR to decode.} The plots show that our low-power envelope detector platform can compute the angle of the AP up to 80~m from the AP {with four antennas}. This range could, in future designs, be improved by increasing the sweep time at AP which in our implementation is 50~ms. Also, the angular error improves with the number of antennas at the AP. Further, with four antennas the range of the design increases because of antenna diversity gains.

\begin{figure}[t!]
\begin{subfigure}{0.35\columnwidth}
    	\centering
    	\includegraphics[width=1\linewidth,height=2.8cm]{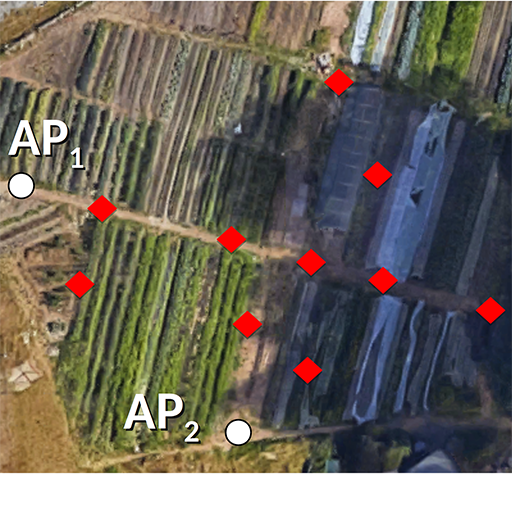}
	\end{subfigure}
    \begin{subfigure}{0.6\columnwidth}
    	\centering
    	\includegraphics[width=1.05\linewidth]{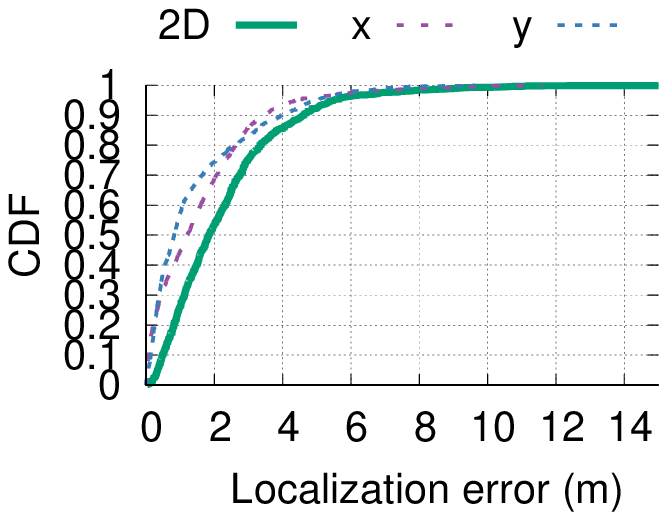}
	\end{subfigure}
  \vskip -1em
  \caption{\small{{2D accuracy with  deployment in farm.}}}
  \label{fig:farm}
  \vskip -0.05in
\end{figure}

{\bf Farm deployment.} We placed one 4-antenna AP at the center of two perpendicular edges of an $90 \times 120$~m farm. We then place the \name\ platform at multiple locations around the farm. We repeat this experiment multiple times at each location and compute the x and y axis errors as well as the 2D location by combining the angles from the two APs. Fig.~\ref{fig:farm} plots the ground truth locations and the CDF of the error at all locations. The plots show that the median 2D localization error is 1.9~m. For context, while prior AoA work  localizes devices at sub-meter resolutions~\cite{robotkyle}, the range at which the experiments were conducted is limited to less than 10~m. As the distance increases, as expected, the same angular error results in a larger localization errors. This meter-scale resolution however is sufficient for our smart farm application.


\subsubsection{High-speed experiments.}  Since we cannot control the bee motion it is difficult to run systematic accuracy experiments at different speeds on an actual bee since we do not know the ground truth at long ranges in outdoor environments. Instead, we place our platform on a DJI Phantom 3 drone. We place the first 4-antenna AP at one end of the soccer field and the second 4-antenna AP as shown  in Fig.~\ref{fig:1drange}. We run experiments at three different speeds using the Phantom 3's wireless controller. We then use the drone's flight record which contains its GPS, altitude, accelerometer and compass data as the ground truth. 
Fig.~\ref{fig:drone} plots the angular accuracy at three different speeds. We plot the results for both the APs. The average angular accuracy at different distances are similar to the stationary angular accuracies observed in the previous experiment. {The fluctuations in error across all distances are due to the changing multipath over time, however the exponential smoothing technique offsets this making some results better than the stationary scenario.}

\begin{figure}[t!]
\begin{subfigure}{0.35\columnwidth}
    	\centering
    	\includegraphics[width=1\linewidth,height=2.8cm]{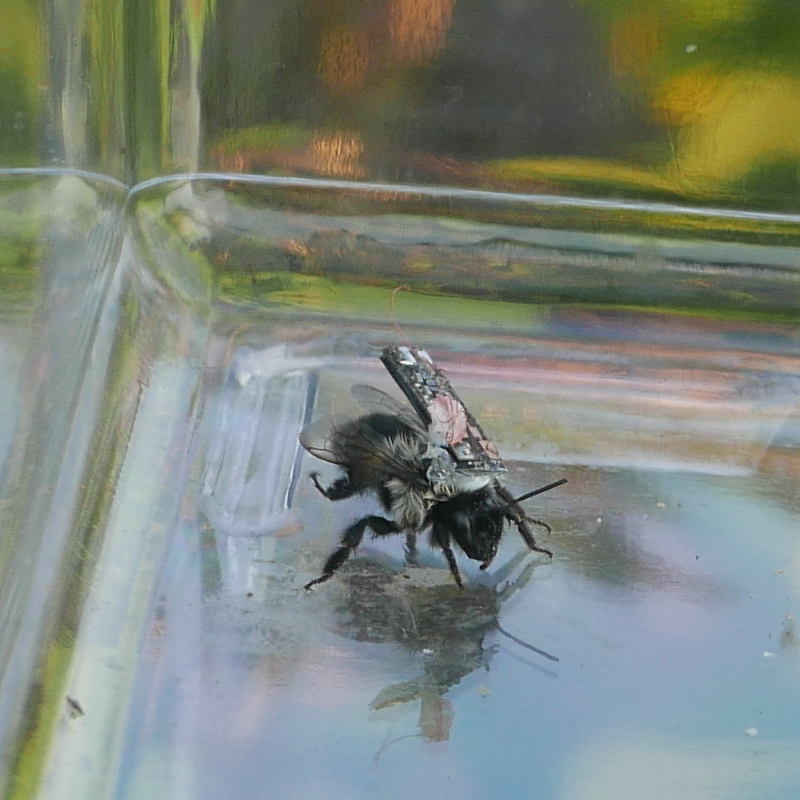}
        \end{subfigure}
    \begin{subfigure}{0.6\columnwidth}
    	\centering
    	\includegraphics[width=1.05\linewidth]{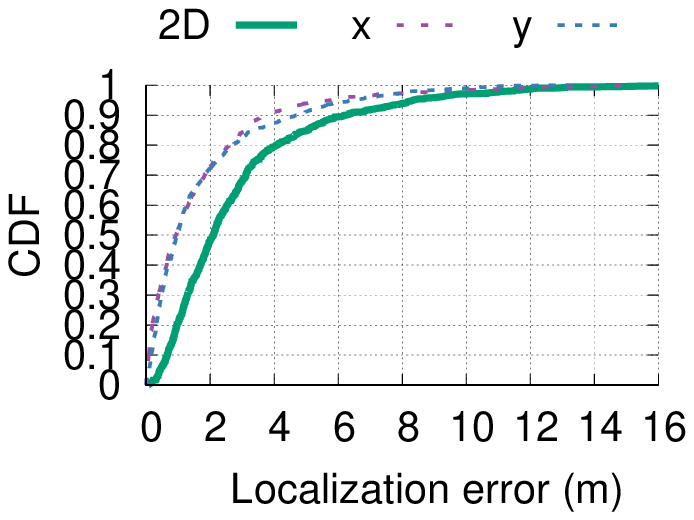}
	\end{subfigure}
  \vskip -1em
  \caption{\small{{Localization of {live bee in enclosure} on the farm.}}}
  \vskip -0.05in
  \label{fig:beecdf}
\end{figure}

\subsubsection{Wild bees observations.}
Over the course of our outdoor experiments we observe many wild bees in both the soccer field and the farm. In total we observe 3 species including honeybees (\textit{Apis mellifera}) as well as two species of bumblebees (\textit{Bombus vosnesenski} and \textit{Bombus sitkensis}). In areas of the farm with flowers in bloom, we observe as many as 20 or more in a single 1~$m^2$ area. Even on the open field we observe many wild bees foraging in weeds with small flowers (e.g., dandelions). We note that while they typically fly at average speeds of 1-5~m/s when going longer distances~\cite{speed1,speed2}, they settle in one region slowly hovering or landing at individual flowers.

\subsubsection{Experiments with bees.} Next, we evaluate our self localization algorithm with our living IoT platform placed on a {live bumblebee as shown in Fig.~\ref{fig:beecdf}. We conduct this experiment with the bee in a plastic container with a volume of $30\times20\times20$~cm, where the bee freely moved around and flew.}  We run experiments again in the farm deployment where we placed the two 4-antenna APs as shown in Fig.~\ref{fig:farm}. We then place the plastic container with the mobile bee at different locations in the farm. The location of the plastic container was taken as the ground truth for computing the localization error. Fig.~\ref{fig:beecdf} plots the CDF of the localization accuracy with our \name\ platform on the bumblebee. This as expected shows that the accuracies are similar to prior experiments and demonstrates the feasibility of self-localization on a live and mobile  bee.

\subsubsection{Effect of flapping wings.}
While a drone serves as an excellent platform to perform systematic experiments with known ground truth data, its flight mechanics are   different from that of an insect. So, we next evaluate the effect of flapping wings on localization accuracy. Unlike propeller driven drones, insects use flapping wings to generate lift. Since the envelope detector is  mounted close to the wings of the insect which move periodically and cause vibration in the body of the insect at the flapping frequency we next evaluate how flapping wings affects the  signal at the envelope detector.

\begin{figure}[t!]
\begin{subfigure}{0.35\columnwidth}
    	\centering
    	\includegraphics[width=1\linewidth,height=2.8cm]{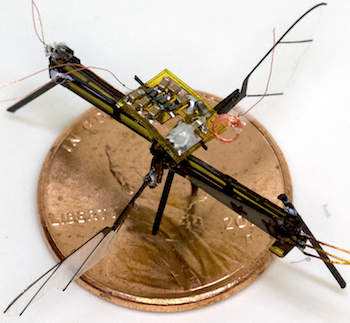}
        \end{subfigure}
    \begin{subfigure}{0.6\columnwidth}
    	\centering
    	\includegraphics[width=1.05\linewidth]{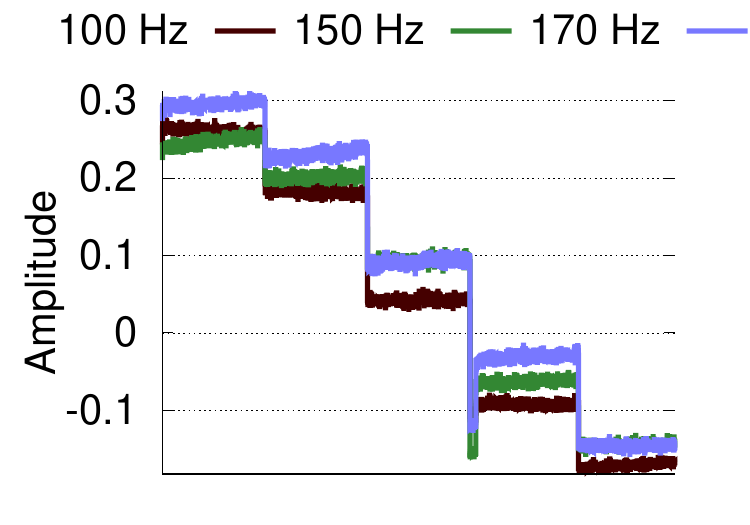}
	\end{subfigure}
  \vskip -1em
  \caption{\small{{Effect of flapping wings on envelope detector.}}}
  \label{fig:flapping}
  \vskip -0.05in
\end{figure}

In order to isolate the effect of flapping from flight motion of a real insect in a systematic manner, we instead attach our envelope detector and antenna to an insect scale robot design with flapping wings that is inspired by~\cite{Wood2008} and shown in Fig.~\ref{fig:flapping}. The robot has a wingspan of 35~mm which is similar to a bumble bee (32~mm) and are designed with hinges to mimic the wing kinematics of real flying insects. 

The robot's wings are driven by 2~piezoelectric bimorph actuators which we control to flap the wings at different frequencies to determine whether the wing motion itself negatively impacts our localization performance. We then incremented the phase difference at the transmitter antennas by discrete steps of thirty degrees and monitored the change in the amplitude of the signal received at the envelope detector. Fig~\ref{fig:flapping} plots the raw received signals at the output of the envelope detector. The figure does not show a noticeable degradation in SNR due to wing motion or vibration of the body as a whole. Further, the different amplitudes created by the phase changes at the AP, appear intact at the output of the envelope detector and are largely independent of the flapping frequency of the insect's wings. 

\subsection{Backscatter Evaluation}

We implement our access point using two USRP software radios. For the transmitter, we configure a USRP to transmit a continuous tone at 915~MHz and connect its output to a Qorvo RF5110G power amplifier which outputs to a 6~dBi patch antenna. We use the same model of patch antenna for our receiver and place it parallel to the transmitter antenna separated by a distance of 0.5~m. We set the receiving USRP to a frequency 2~MHz away from the transmitter corresponding to the subcarrier generated by backscatter. We then apply a band pass filter in software to remove unwanted noise.

\begin{figure}[t!]
\begin{subfigure}{0.48\columnwidth}
    	\centering
    	\includegraphics[width=1.1\linewidth]{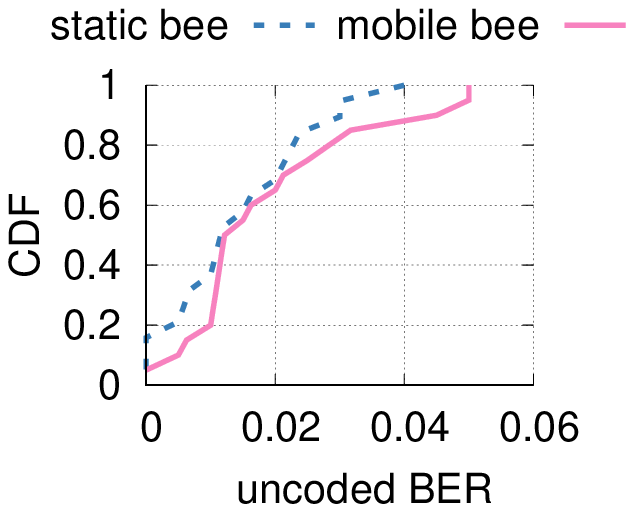}
        
	\end{subfigure}
    \begin{subfigure}{0.48\columnwidth}
    	\centering
    	\includegraphics[width=1.1\linewidth]{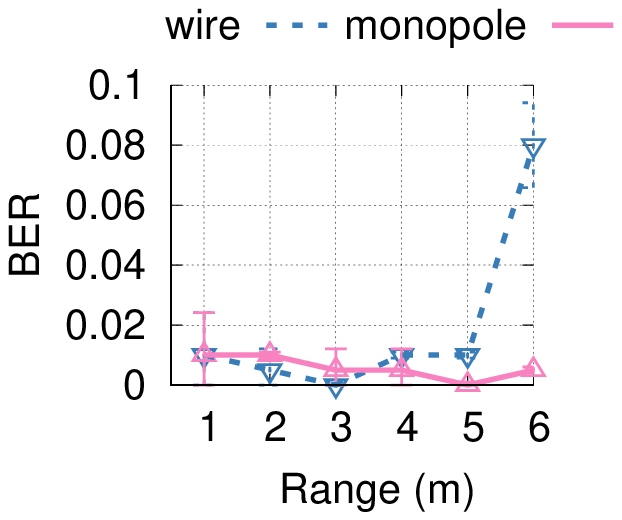}
       
	\end{subfigure}
  \vskip -1em
  \caption{\small{{Backscatter  performance.} The uncoded BER is low and comparable to prior backscatter designs~\cite{fmbackscatter} and the bee can upload data when it is back at the hive.}}
  \label{fig:ber}
  \vskip -0.05in
\end{figure}

We evaluate our setup in our application scenario of the bee uploading its data when it is near the hive. We evaluate this first with a dead bee placed at fixed locations, and then on a live bee in a plastic enclosure. We ensure that our evaluation board is light enough to enable the bee to fly and walk around the enclosure; we glue the board to the top of the insect's abdomen. We place our AP approximately 1~m from a bee hive separated from the AP by an enclosure consisting of 2 layers of insect netting. We configure the AP to continuously transmit 100~bits and measure the error rate at 20~locations within 2~m of the hive. 
We place the hive on a 1~m high table and include points on every side of the hive to simulate bees taking a variety of trajectories to approach the hive. We plot the uncoded BER in Fig~\ref{fig:ber}. The plot shows that the live bee has a similar performance to that of the dead bee demonstrating that our backscatter system can operate in the presence of bee motion. We note that the uncoded BERs are similar to prior backscatter works~\cite{fmbackscatter} and  can be further reduced by applying  error correcting codes~\cite{abc,lorabackscatter}.

Finally, to understand the effect of the antenna without complicating factors of a live insect and its motion, we evaluate in a stationary case, with both our lightweight wire antenna as well as a standard monopole antenna.  We place our AP at a fixed location in  an outdoor environment and configure our backscatter device to transmit a known sequence of 200~bits at a rate of 1~kbps. We then move it to increasing distances  away from the AP and record the received bits at each location. Fig.~\ref{fig:ber} plots the uncoded BER versus distances  and show that our light-weight antenna can achieve  low BERs upto 5~meters. Beyond 5~m, the monopole antenna performs better than the light-weight antenna. This 5~m range is however sufficient for our application where the bees upload the data when they are back at the hive.

\subsection{Power harvesting Feasibility}

\subsubsection{Recharging batteries using RF} 
\begin{figure}[t!]
\begin{subfigure}{0.35\columnwidth}
    	\centering
    	\includegraphics[width=1\linewidth,height=2.8cm]{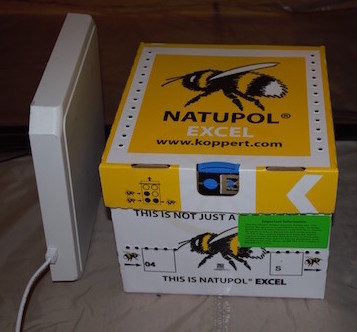}
	\end{subfigure}
    \begin{subfigure}{0.6\columnwidth}
    	\centering
    	\includegraphics[width=1.05\linewidth]{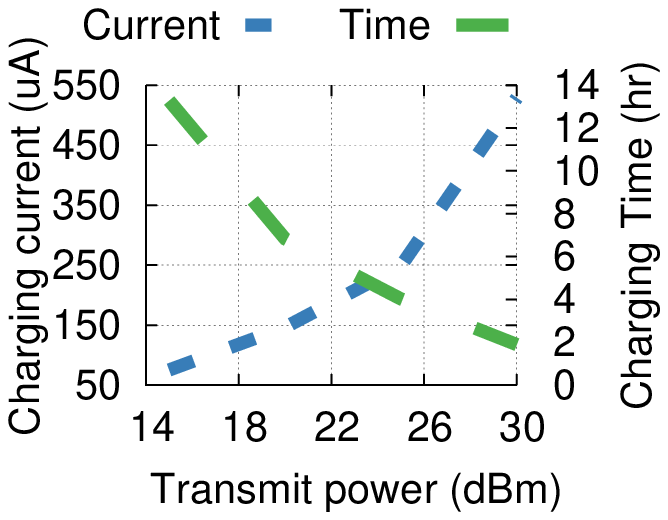}
	\end{subfigure}
  \vskip -1em
  \caption{\small{{Recharging batteries using RF harvesting.}}}
  \vskip -0.05in
  \label{fig:rf}
\end{figure}

Since certain insects like bees return to a single colony or hive we can use this as a central charging point. To evaluate the feasibility of this approach, we first measure the attenuation of 900~MHz signals through a real bumble bee hive (Natupol, Koppert)~\cite{koppert}. We place a 900~MHz 6~dBi patch antenna AP in different locations near the hive including above, below, and on either side. We then place our platform on the opposite side. We find that placement of the transmit AP below the hive causes significant attenuation as these commercial hives contain a package of sugar water below the hive to provide a food source. We also find that placing the AP above the hive results in the lowest path loss of approximately 15~dB, while the side and back are 3 and 6~dB worse respectively. Fig.~\ref{fig:rf} plots the measured harvested power  versus transmitted RF power from the top of the hive.  The results show that a 20~dBm AP could charge a 1~mAh battery in about 6~hours. 

\subsubsection{Solar harvesting for battery-free platform.}\label{sec:solar} 
\begin{figure}[t!]
\begin{subfigure}{0.35\columnwidth}
    	\centering
    	\includegraphics[width=1\linewidth]{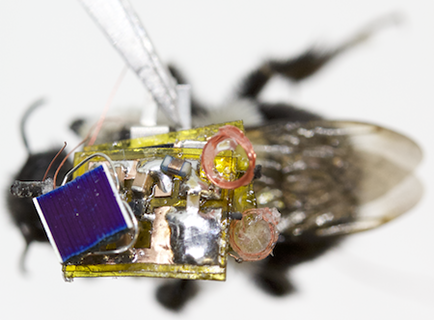}
	\end{subfigure}
    \begin{subfigure}{0.6\columnwidth}
    	\centering
    	\includegraphics[width=1.05\linewidth]{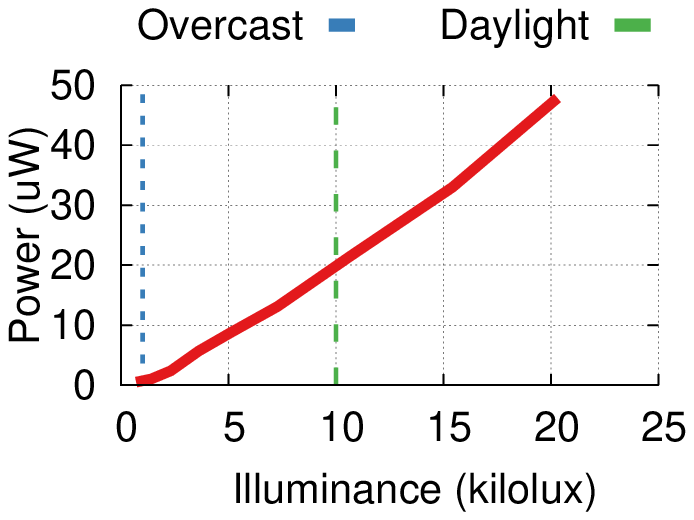}
	\end{subfigure}
  \vskip -1em
  \caption{\small{{Solar harvesting for battery-free platform.}}}
  \vskip -0.15in
  \label{fig:solar}
\end{figure}

Insects such as bumblebees are most active during daylight hours. This makes solar power harvesting a particularly attractive option as these creatures naturally fly in outdoor environments with abundant sunlight. We evaluate an 8~mg 3x3~mm photovoltaic cell~\cite{mhgopower} shown in Fig.~\ref{fig:solar}(a) by measuring the output of the cell under a microscope light~\cite{amscope} with controllable intensity. We measure the output current and voltage of the cell  and plot the results for illuminance values ranging from 1000-20,000 Lux in Fig.~\ref{fig:solar}(b). At 1000~Lux which is representative of an overcast day the PV cell can harvest 1~$\mu$W of power and up to 50~$\mu$W on clear sunny days. This shows the potential for a battery-free design that could replace the 70~mg battery with a solar cell and small storage capacitor.

\subsection{Sensor Peripherals}
\begin{figure}[t!]
  \centering
  \includegraphics[width=0.4\linewidth]{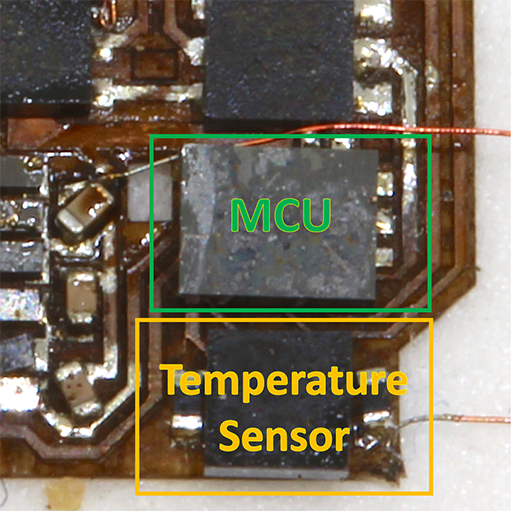}
  \hspace{2em}
  \includegraphics[width=0.4\linewidth]{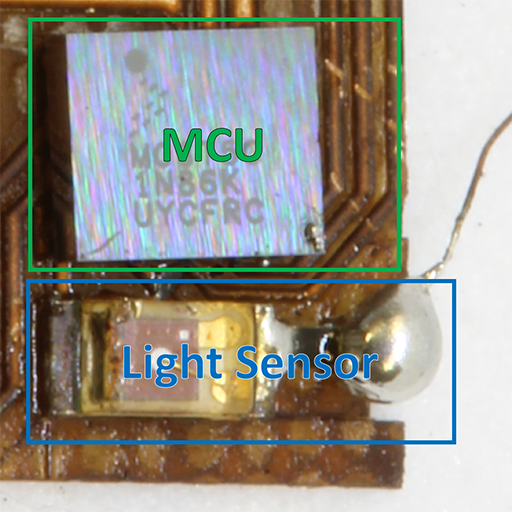}
  \vskip -1em
  \caption{\small{Close up images of temperature and humidity sensor (left) and ambient light sensor (right).}}
  \label{fig:sensor}
  \vskip -0.05in
\end{figure}

We integrate three sensors: temperature, humidity, and light intensity. These sensors are  commercially available in small packages compatible with our weight budget and operate at low power.
For performing temperature and humidity measurements we use the TI HDC2010 IC which weighs  3~mg and includes the physical sensor as well as an integrated ADC and digital interface as shown in fig.~\ref{fig:sensor}. This chip is capable of providing high accuracy measurements while consuming as little as 0.55~$\mu$A of current for measuring both sensors once per second. Additionally, its low power sleep mode allows for greater power savings by leaving it inactive until our  algorithm detects we are in a location we wish to sense. We also measure light intensity using the ALS PT19~ambient light sensor. This chip weighs only 1.5~mg, but does not include integrated readout electronics. Instead, we use the built in ADC on our microcontroller to periodically sample the analog output.

During operation, \name\ must also be able to log its sensor data until returning to the hive to upload it. The humidity and temperature values require 11~bits each, while the light intensity requires 12~bits. Our microcontroller has up to 32~KB of onboard flash memory. Using 2 bytes for each measurement, 1 byte for each angle, is a total of 4 bytes for a sensor measurement and 2 angle values. This would allow us to store measurements once every 5s for over 10~hrs. This is sufficient to cover daylight hours during which bees are active and foraging. Alternatively, we can leverage our localization technique to selectively log data at higher rates near specific target locations.

\section{Related Work}
\label{sec:related}

{\bf Tracking bees.} Prior biological research has explored the problem of tracking bees to understand their behavior and help explain the decline in their worldwide population.~\cite{lasertrack} attaches bees with  laser light-activated microtransponders  which can be detected by laser readers at distances of up to 10 mm from the reader. To know when the bees enter and leave the hive, a 10 by 10 mm plastic tube walkway was attached to the hive entrance, with two laser readers at the top of the tube. 
Intel and the Australian Commonwealth Scientific and Industrial Research Organization (CSIRO) are exploring the possibility of attaching RFID tags on bees to track them at a range of less than 10~cm allowing them to notice when the bees enter and leave the hive~\cite{beeart1,beeart2}. Backscatter  has also been used with large dragonflies~\cite{mattreynolds}; it uses a custom silicon design and also does not have RF localization capabilities.

Radar based systems~\cite{naturepaper1999,hornet} use a pulse radar with a  25~KW transmit
power at 9~GHz as the reader and use a passive analog diode on an insect that creates harmonics at 18~GHz which the reader uses to track the insect at much longer distances~\cite{naturepaper1999}.  The key challenges with this design are three-fold: 1) the analog design with the diode always creates a 18~GHz harmonic and hence cannot support more than one bee, 2) these passive analog diodes do not support general-purpose sensors  on the bee, and 3) the bee cannot localize itself and has no computing capabilities.

Finally,~\cite{beeradio} uses a 200~mg active pulse radar transmitter at 100-200~MHz powered by a small battery on a large species of European bumblebee. These transmitters send small pulses which the reader uses to track the bee. This does not satisfy our requirements for five reasons: 1) across our experiments the bees common in our environment, were unable to lift more than 105~mg, 2) the bees do not have the localization information and so it does not support self-localization, 3) it is unclear if the design will scale to large numbers of bees since they all must transmit the corresponding pulses, requiring a scalable MAC protocol, 4) transmitting  high power pulses at the bee  would limit other operations that can be performed and  5) it does not carry general-purpose sensors.

In contrast to this prior work, we make the following contributions: i) design the first low-power self-localization technique for flying insects, ii) present a general purpose platform that enables computing, communication and sensing on aerial insects and, iii) demonstrate for the first time that insects such as bees can be used to carry general purpose sensors and thus enable mobility, in lieu of drones.

\vskip 0.05in\noindent{\textbf{Controlling insect flight.} Researchers have shown how to control the motion of larger insects such as beetles, dragonflies and Locusts~\cite{dragoncontrol, michel_beetle, michel_locust,additionalcontrol}. An interesting future research direction would be to develop low power flight control for small insects like bees. This however is challenging since while the nervous system of larger insects like beetles is well understood, bees are at least an order of magnitude smaller in size and it has not been shown that their flight can be controlled. Achieving this for small insects like bees is an open problem and, if solved, can augment our work.}

\vskip 0.05in\noindent{\bf Power harvesting from large insects.} Prior work has tried to harvest power from large insects such as moths and beetles~\cite{harvest1}.~\cite{harvest2, harvest3} utilizes a piezoelectric transducer which converts the vibratory 
motion created  by the insect's flight into electrical power to harvest 7--60~uW from beetles and hawk moths.~\cite{harvest4} harvests 0.8~mW from moth vibrations using a magnetic induction generator. Finally,~\cite{harvest6}  harvests around 1~$\mu$W from the chemical energy stored within a moth's hemolymph. These systems however use moths and beetles that are more than 10 times larger than bees. 
Instead we show RF harvesting to recharge the battery and the feasibility of using solar cells to enable battery-free designs.

\vskip 0.05in\noindent{\bf Bio-inspired aerial robots.} The robotic community has spent the last two decades in the design of insect-scale aerial robots~\cite{hardware1,hardware2,fearing2000} that are similar in size to houseflies and mimic the wing propulsion of insects. Despite significant research~\cite{Brooks1989a, Wood2008,ma2013controlled,ocelli,Graule2016}, these robots are largely tethered to a wire to power and control them, since they consume hundreds of milliwatts of power for the mechanical propulsion and cannot carry batteries given the weight requirements. While recent work shows the feasibility of wireless power using lasers at the range of a meter~\cite{robofly}, providing hundreds of milliwatts of power at 80--100~m is challenging and difficult to scale beyond a single aerial robot. Our idea instead is to leverage biological insects which can be thought of as efficient biological machines that provide flight and piggyback communication, computing and sensing on top of them.

\vskip 0.05in\noindent{\bf Size-constrained sensor systems.} \cite{michigan1,michigan2,michigan3} designed die-stacked sensor platforms that have the key building blocks of miniaturized sensor nodes, such as data transceivers, energy harvesters, power management units, and digital logic circuits. {This approach requires custom ICs which limits availability to researchers elsewhere as well as programmability. We take an alternate approach by designing a  system using commercial off the shelf components based on a microcontroller that is programmable to create a more modular platform.} In addition, we also  demonstrate a self-localization technique using only an envelope detector.

{~\cite{msr_gps, patwari2005, xi2010locating} design localization systems for sensor networks with a focus on decreasing cost, power, or improving accuracy; however these techniques all require active radios and cannot easily be scaled down to the size and power requirements for use on insects like bees.}

Our prior work~\cite{sensys2018} designs a sub-centimeter sized backscatter device based on a microcontroller and localizes it using LoRa backscatter \cite{lorabackscatter,netscatter} at a software radio using non-linear optimization on I/Q samples. In addition to being focused on a novel  mobile insect application, our design differs from~\cite{sensys2018} in four key ways: First, our localization algorithm is not based on backscatter and does not occur at the software radio. Instead we design a localization algorithm that runs on the low-power device using the output of an envelope detector, without access to the I/Q samples. Second, our backscatter design does not use LoRa transmissions, which significantly simplifies our design. Third, a key limitation of the design in~\cite{sensys2018} is that it cannot  currently scale to more than one to two devices. In contrast, our self-localization algorithm is similar in spirit to GPS and hence can scale to a large number of insects at the same time. Fourth, unlike non-linear optimizations that cannot effectively run on our low-power microcontroller, we design a low-complexity algorithm to estimate the 2D location at the Living IoT platform.

\section{Discussion and Conclusion}
\label{sec:discussion}
This paper explores  the use of insects in lieu of drones to enable mobility for sensor networks. 
Making this vision pervasive, however, requires addressing additional challenges. 
\squishlist

\item {\bf Weather dependency.} Bees hibernate during winter in cooler climates. This however correlates with plant growth and increased activity on farms in warmer seasons. 

\item {\bf E-waste.} Piggybacking on insects could lead to electronic waste being scattered around the farms when the insects eventually live out their lifespan. There are three approaches to addressing this problem: i) ensure that the electronics are removed before the expected lifetime of the bees, ii) the electronics can be localized for a while even after the bees die which can be used for cleanup purposes and  iii) use biodegradable electronics~\cite{biodegradable1,biodegradable2} in the design of \name.

\item {\bf Fabrication.} {{Our current prototype requires manual fabrication and attachment to the insects. However  we use commercially available parts which allows for easy scaling of the electronics fabrication} and our process for gluing the electronics to the insects is similar to the process for attaching tracking markers to bees in commercial hives. Additionally, researchers have also shown that insects can survive common microfabrication processes such as deposition of conductive material in a vacuum chamber~\cite{fly_mems1,fly_mems2} and performing surgeries at different stages of an insect's life cycle. This suggests potential approaches for mass attachment of electronics or fabricating devices on insects themselves. Further work on implanting devices at different stages of an insect's life cycle also has potential to improve fabrication~\cite{pupa_surgery}.}

\item {\bf Camera sensing.} A future research direction is to integrate cameras~\cite{cameransdi} with the Living IoT platform. This can be useful for smart farm applications like canopy monitoring.  Centeye image sensors and cameras such as the Himax HM01B0 offer a potential path for achieving such a camera-based sensing system within our weight/power budget.

\item{{\bf IACUC requirements.} Finally, we note that while working with insects and other invertebrates is not governed by the Institutional Animal Care and Use Committee (IACUC) policies of our institution and insect consciousness is poorly understood~\cite{merker2016insects}, we do our best to follow the three Rs of animal research~\cite{three_rs}. We minimize use of insects by benchmarking each aspect of our system using drones to simulate flight and robots to simulate flapping wings. For our experiments we perform no surgical modification, we simply attach weight to the exoskeleton and we observe no significant changes in behavior after the procedure. Additionally we use only a small number of insects for our experiments and remove our electronics package after  completing experiments. We also release wild caught insects back in the area they were captured after completing experiments.}

\squishend

\section{Acknowledgments} {{We thank Swarun Kumar, Tom Daniel, Callin Switzer, members of the Air Force center of excellence on Nature Inspired Flight Technologies and Ideas (NIFTI), members of the UW Networks and Mobile Systems group as well as the anonymous reviewers at MobiCom for their constructive feedback on this work. The authors are funded in part by NSF awards CNS-1812554, CNS-1452494, CNS-1823148, Google Faculty Research Awards and a Sloan Fellowship.}}

\bibliographystyle{acm}

\balance{}
\bibliography{reference}
\end{document}